# Statistical properties and shell analysis in random cellular structures

T. Aste§, K. Y. Szeto, and W. Y. Tam

Department of Physics

The Hong Kong University of Science and Technology

Clear Water Bay, Kowloon, Hong Kong.

§Laboratorie the Physique Théorique,

Université Louis Pasteur, Strasbourg France

December 14, 1996

## Abstract

*We investigate the statistical properties of two dimensional random cellular systems (froths) in term of their shell structure. The froth is analyzed as a system of concentric layers of cells around a given central cell. We derive exact analytical relations for the topological properties of the sets of cells belonging to these layers. Experimental observations of the shell structure of two-dimensional soap froth are made and compared with the results on two kinds of Voronoi constructions. It is found that there are specific differences between soap froths and purely geometrical constructions. In particular these systems differ in the topological charge of clusters as a function of shell number, in the asymptotic values of defect concentrations, and in the number of cells in a given layer. We derive approximate expressions with no free parameters which correctly explain these different behaviors.*

## 1   Introduction

Materials consisting of cellular structures such as metal grains and biological tissues are common in nature [1, 2]. Among these systems, soap froth is considered as the paradigm for the study of trivalent two-dimensional cellular structures. The structural analysis of two-dimensional cellular patterns have been made by many researchers and many interesting results have been obtained [3, 4, 5, 6, 7, 8, 9, 10, 11, 12]. These studies of the topological properties of soap froth can be summarized in several laws such as the Lewis' law[13] on the the statistics of cell area, the von Neumann's law[14] on the growth rate of $n$-sided cells, the scaling law[6] on the probability distribution of the cells, and the Aboav-Weaire's law on nearest neighbor correlation [15, 16]. So far, the evolution of soap froth after the scaling state defined as stationary probability distribution of cell sides have been explained by several theories [17, 18, 19, 20], indicating that correlation effects are not manifested in the analysis of area scaling law. However, more detailed analysis beyond area scaling law has been done[21] and suggests strongly the importance of clarifying the role of correlation effects. Moreover, there has never been any experimental investigation till recently[22] on the statistical properties and correlation effects beyond nearest neighbors in two-dimensional soap froth. With the advance of experimental techniques and data analysis[23, 24, 22], it is natural to investigate the structural characteristic of soap froth beyond nearest neighbors. Here we derive systematically exact expressions for the topology of the froth structure beyond first neighbours. Our results will serve as a mathematical framework for our data analysis, which enhances our ability to test the current theoretical ideas and our understanding of topological ordering processes in soap forth [25, 26, 27, 28, 29]. Our aim is to extract more information on the differences between physical soap froths and computer generated two-dimensional cellular patterns, such as the Voronoi construction of points randomly scattered on the plane.

Soap froths are trivalent (three edges meeting at a vertex) two-dimensional cellular pattern. Any two-dimensional ($2D$) froth can be analyzed as structured in concentric layers of cells around a given central "germ" cell[25, 30, 22]. The first layer of the shell structure consists of the neighbours of the germ cell, the cells of the second layer are the neighbours that are bounding externally the first layer, etc. (see Fig.(1)). More precisely, the layers are closed rings of cells which are at the same topological distance from the germ cell (the topological distance between two cells is defined as the minimum number of edges that a path must cross to connect these two cells). The shell is closed loop of edges which are the interface between two subsequent layers (see Fig.(1)). (The definition here is consistent with ref.[30]. Note that in our previous work[22], the word shell is used to indicate the layer used in this paper. )

The shell structure has some physical relevance in natural froths and in problems of topologically-stable partition of space by cells. In foams the evolution of the cellular structure is ruled by the process of diffusion of the gas inside a given bubble towards its neighbours. The topological distance $j$ between two cells is the minimum number of soap membranes that the gas molecules have to cross to pass from one cell to the other. The molecules which diffuse from a given central bubble move through the shell structure reaching the cells of each layer with approximately equal probability. An analogous problem where the shell structure has a physical relevance is the random walk through a barrier network. In this case the topological distance $j$ is the minimum number of barriers that must be crossed from a starting-cell to a given final cell. The number of cells in the layer $j$ is the number of final states with approximately equal probabilities. More generally, any perturbation on a given cell (cell growth, cell division, mechanical stress, cell-coalescence, electrical signal...) propagates in the whole system through the shell network and reaches with approximately equal intensity to all cells at the same topological distance from the perturbed cell at about the same time. These considerations provide the motivations of introducing shell structure analysis for froth.

The quantities investigated for a shell at topological distance $j$ from a given $n$-sided germ cell are: the total number of cells in the layer ($K_j$), the average number of sides of the cells belonging to the layer ($m_j$), and the topological charge ($Q_j$) of the cluster of cells delimited by the shell $j$. Here, the topological charge of a cell with $n$ sides is defined as $q = 6 - n$ and $Q_j$ is defined as the sum over the topological charges of all the cells inside the cluster delimited by the shell $j$ which separates the layer $j$ from the layer $j + 1$. Note that all these quantities, $K_j$, $m_j$, and $Q_j$ can depend on $n$.

The experimental investigation of the statistical properties of natural and computer generated froths in term of the shell structure give some interesting results. For instance, we find that the number of cells in a layer at a topological distance $j$ from the germ cell increases with the distance following a linear law $K_j = Cj + B$ with slope $C \sim 9$ (see also Ref.[22]). This is in contrast to simple geometrical considerations that the perimeter of the shell cluster increases with the radius with a slope approximately $2\pi$. Moreover, we find that the topological charge of the cluster bounded by the shell $j$ is negative and decreases linearly with $j$. This behaviour is particularly surprising since the total topological charge of a froth is a constant finite quantity independent of the network itself and is related to the space curvature by the Euler formula associated with the Gauss-Bonnet theorem[31, 32]. Therefore, the set of cells belonging to a layer have peculiar statistical properties which are different from the one of the whole froth. The aim of the present paper is to study these peculiarities with an exact analytical approach, and to find approximate solutions with no adjustable parameters which can be compared with the experimental observations.

The plan of the paper is as follow. In Section 2, we derive the statistical properties of the shell structure for a special class of froths called shell-structured-inflatable (SSI) [30]. These froths are particularly convenient since they can be constructed layer after layer in a recursive way according to an inflationary procedure. In Section 3, we study the correlation length in SSI Euclidean froths. Approximate expressions for $K_j$ and $Q_j$ are also derived. In Section 4, the results obtained for SSI froths are generalized to non-SSI systems. In section 5, we find approximated expressions for $K_j$ and $Q_j$ in the general case of non-SSI froths. In Section 6, experimental results are presented and compared with the analytical predictions. A conclusion emphasizes the main results. In Appendix A, a generalization of the Weaire's sum rule [16] is derived. Finally, in Appendix B, we find an expression for the fluctuations of the topological charge.

# 2    Statistical properties in Shell-Structured-Inflatable Froths

Any froth can be analysed as structured in concentric layers of cells which are at the same topological distance from a given central cell. These concentric layers are the shell structure of the froth. In this structure the cells making the layer $j$ can be put into two categories. Some cells have simultaneously neighbours in the layers $j-1$ and $j+1$. These cells make themselves closed layers and constitute the "skeleton" ($sk$) of the shell-structure. Other cells (or clusters of cells) are inclusions between the layers of the shell-skeleton (they have neighbours in the layer $j-1$ or other topological inclusions but not in the layer $j+1$). The shell-skeleton is itself a space-filling froth hierarchically organized around the germ cell. Once a germ cell is chosen, the shell-structure and its skeleton are univocally defined, but different germ cells generate different skeletons. We call a froth shell-structured-inflatable (SSI) if it is free of topological inclusions. (In this paper these inclusions are also called topological defects.) For SSI froths the shell-structure and its skeleton coincide.

In this section we analyzed only SSI systems (no topological defects), the results obtained for such class of systems are extendible to froths with topological defects. The advantage of studying SSI froths is that in these systems the shell-structure can be constructed by using an inflationary recursive process. In particular, one finds that the number of vertices in successive shells are related by the following map[30],

$$\left( \begin{array}{c} V_j^+ \\ V_j^- \end{array} \right) = \left( \begin{array}{cc} s_j & -1 \\ 1 & 0 \end{array} \right) \left( \begin{array}{c} V_{j-1}^+ \\ V_{j-1}^- \end{array} \right) \quad , \tag{1}$$

where $s_j = m_j - 4$, and $m_j$ is the average number of sides per cells in layer $j$. The quantity $V_j^+$ ($V_j^-$) is the number of vertices attached to edges directed outward from shell $j$ (directed inward to shell $j$) (see Fig.(1)). The matrix equation (1) is the logistic map [34]. It is a dynamical map from the central germ cell ($j = 0$) to the whole froth.

Since our $2D$ froth is trivalent (three edges meeting at one vertex), the number of cells in layer $j$ is $K_j = V_j^-$. In term of these quantities eq.(1) can be written as a recursive equation

$$K_{j+1} = s_j K_j - K_{j-1} \quad . \tag{2}$$

The initial conditions are $V_0^- = K_0 = 0$ and $V_0^+ = K_1 = n =$ number of neighbours of the central germ cell.

Given the set of parameters $\{s_j\}$, the solutions of the eq.(1) or (2) are particular trajectories in the plane $(j, K_j)$. In the case $s_j$ equals to a constant $s$, one finds that the parameter $s$ separates the map into different classes.[30] Values $|s| < 2$ are associated with the elliptic region which has bounded, finite trajectories. The region $s > 2$ is the hyperbolic region associated with exponential trajectories. The point $s = 2$ divides the elliptic from the hyperbolic region and corresponds to tilings of the Euclidean plane. In this case one finds that the solutions are linear trajectories $K_j \propto j$. In general, when $s$ depends on $j$, one can note that bounded trajectories always correspond to froths which are tiling elliptic manifolds whereas unbounded trajectories which grow faster than $j$ correspond to tilings in hyperbolic manifolds. The Euclidean space has trajectories between these solutions associated with spaces of opposite curvatures. They are unbounded trajectories which grows asymptotically as a linear law ($K_j \propto j$).

One finds that the total topological charge of a cluster of cells bounded by the shell $j$ can be written in terms of numbers of vertices coming in and going out from the shell $j$

$$Q_j = \sum_i (6 - n_i) = 6 - V_j^+ + V_j^- \tag{3}$$

where the sum runs over all the cells $i$ in the cluster and $n_i$ is the number of sides of the cell $i$[35]. Equation (3) is a general expression (valid also for non SSI systems). It states that the total topological charge inside a cluster depends only on its boundary. For SSI froths eq.(3) can be rewritten as

$$Q_j = 6 - K_{j+1} + K_j \quad . \tag{4}$$

The inverse of this equation provides a relation between the number of cells in the layer $j$ and the topological charge of the cells inside the cluster delimited by the shell $j$

$$K_j = 6j - \sum_{i=0}^{j-1} Q_i = 6j - 6 + n - \sum_{i=1}^{j-1} Q_i \quad , \tag{5}$$

where we define $Q_0 = 6 - n$. Let us now make use of these recursive relations in order to evaluate the quantities $K_j$, $Q_j$, $m_j$, in the shell structure.

## 2.1 First shell: the Aboav-Weaire's law

Consider a shell-structure around an $n$-sided germ cell. The number of cells constituting the first layer ($j = 1$) is

$$K_1 = n \quad , \tag{6}$$

the topological charge inside the cluster delimited by the first shell is by definition

$$Q_1 = (6 - n) + (6 - m_1)K_1 \quad . \tag{7}$$

Using the sum rule (46) (see appendix A), one obtains the average value

$$\langle Q_1 \rangle = -\mu_2 \quad , \tag{8}$$

with $\mu_2 = \langle (n-6)^2 \rangle$. Here the averages $\langle (...) \rangle$ are over the cell-sides distribution: $\langle (...) \rangle = \sum_n p(n)(...)$, with $p(n)$ being the probability of an $n$-sided cell in the whole froth. One can introduce the fluctuation part of topological charge $Q_1$ as its deviation from the average value: $\Gamma_1 \equiv Q_1 - \langle Q_1 \rangle$, where a-priori the fluctuation part can be any possible function of $n$ satisfying the condition $\langle \Gamma_1 \rangle = 0$. In terms of the fluctuation, the charge $Q_1$ can be written as

$$Q_1 = (6 - n) + (6 - m_1)n = -\mu_2 + \Gamma_1 \quad . \tag{9}$$

Equation (9) gives the relationship between the number of sides ($n$) of a given germ cell and the average number of sides ($m_1$) of the cells neighbouring this $n$-sided cell. In literature, such a relation is widely studied since Aboav founds empirically the linear law: $(m_1 - 6)n = \mu_2 - a(n - 6)$ [15]. One can see immediately that the Aboav-Weaire's law is obtainable from eq.(9) by imposing a linear form for the fluctuation part (i.e. $\Gamma_1 = \epsilon_1(6 - n)$ with $\epsilon_1 = 1 - a$). This linear dependence can be interpreted in terms of the screening of the central charge $Q_0 = 6 - n$ by the charges of the first layer. The total screening charge in the first layer is $(6 - m_1)n$ and its deviation from the average is

$$(6 - m_1)n - \langle (6 - m_1)n \rangle = -a(6 - n) \quad . \tag{10}$$

We can therefore interpretate $a$ as a screening factor: $a = 1$ corresponds to a total screening of the internal charge, whereas $a = 0$ corresponds to absence of screening.

The Aboav-Weaire's law is generally associated with the presence of topological correlations between nearest neighbours cells. In Appendix B we discuss in details the effects of finite range correlations in froths. Let's note that an arrangement of cells free of correlation (called topological gas [36]) has $m_1 = 6 + \mu_2/6$ [37], which leads to an Aboav coefficient $a = -\mu_2/6$ and $m_1$ is independent of n.

## 2.2 Generic $j$-shell

The average topological charge inside the cluster delimited by shell $j$ is

$$\langle Q_j \rangle = \sum_{i=1}^{j} \langle (6 - m_i)K_i \rangle \quad . \tag{11}$$

Indeed, the topological charge is an additive quantity and therefore $\langle Q_j \rangle$ is the sum of the topological charges contained in the layers $i$ ($\langle (6-m_i)K_i \rangle$) which are inside the cluster with 'radius' $j$ ($i \leq j$). As before, we introduce the fluctuations of the topological charge in a cluster

$$\Gamma_i = Q_i - \langle Q_i \rangle \quad . \tag{12}$$

By using the generalized Weaire identity $\langle (6-m_j)K_j \rangle = \langle (6-n)K_j \rangle$ (eq.(46) in appendix A) and by substituting eq.(5) into (11), we can express the portion of charge contained inside the layer $j$ in term of the topological-charge fluctuations

$$\langle (6-m_j)K_j \rangle = -\mu_2 - \sum_{i=1}^{j-1}\langle (6-n)\Gamma_i \rangle = -\sum_{i=0}^{j-1}\langle (6-n)\Gamma_i \rangle \quad , \tag{13}$$

where we used $\Gamma_0 \equiv (6-n)$. The total charge of the cluster inside the shell $j$ is therefore

$$\langle Q_j \rangle = -\mu_2 j - \sum_{i=1}^{j-1}(j-i)\langle (6-n)\Gamma_i \rangle = -\sum_{i=0}^{j-1}(j-i)\langle (6-n)\Gamma_i \rangle \tag{14}$$

Note that, in the absence of fluctuations ($\Gamma_i = 0$ for $i > 0$), eq.(13) give a topological charge per layer equal to $-\mu_2$ and therefore a total charge which decreases linearly in $j$ with slope $-\mu_2$.

Using eq.(14) and the definition of $\Gamma_j$, one can rewrite eq.(5) in terms of the fluctuations

$$\begin{aligned}
K_j &= 6j - \sum_{i=0}^{j-1} Q_i \\
&= 6j - (6-n) - \sum_{i=1}^{j-1} Q_i \\
&= 6j + n - 6 + \frac{j(j-1)}{2}\mu_2 - \sum_{i=1}^{j-1}\Gamma_i + \sum_{i=1}^{j-2}\frac{(j-i)(j-i-1)}{2}\langle (6-n)\Gamma_i \rangle \\
&= 6j - \sum_{i=0}^{j-1}\Gamma_i + \sum_{i=0}^{j-2}\frac{(j-i)(j-i-1)}{2}\langle (6-n)\Gamma_i \rangle
\end{aligned} \tag{15}$$

These relations for $K_j$, $Q_j$ and $m_j$, are exact results valid for any SSI froth. The fluctuations $\Gamma_j$ are a-priori unknown functions subjected to the constraint $\langle \Gamma_j \rangle = 0$. Other constraints on these fluctuations come from the space-filling condition.

## 3  Shell statistic of Euclidean froths

Let us consider an Euclidean $2D$ froth. In such a froth the number of cells per layer must grow linearly with the distance in the asymptotic limit. We will show that fluctuations in the topological charge ($\Gamma_j \neq 0$) are essential for filling space with disordered ($\mu_2 \neq 0$) Euclidean SSI cellular systems. Consider a system with $\Gamma_i = 0$ for $i \geq 1$ (recall $\Gamma_0 = 6-n$), from the last equality of equation (15), one obtains that the number of cells in the generic layer $j$ increases quadratically with the distance $K_j \propto j^2\mu_2$. Such a froth is realizable only in a space with intrinsic dimension $D = 3$ and thus it is not two-dimensional Euclidean. In natural froths and in computer generated cellular system the shell structure is organized in order to keep the froth Euclidean and experimentally one finds that the number of cells per layer increases linearly in $j$ after the first few layers[22]. Such organization can be provided by the $\Gamma$'s which must be different from zero at least for the first few shells.

We have the following interesting theorem. (See Appendix C for a proof.) For an Euclidean SSI froth with $\mu_2 \neq 0$ that obeys the Aboav-Weaire's law, cells must be correlated at least between the third neighbors. This theorem can be restated into two parts: (1) If an Euclidean SSI froth with nonzero second moment obeys Aboav-Weaire's law, the minimum $\nu$ in order that $\langle K_j \rangle \propto j$ for $j \geq \nu$ is 3. (2) Under the same hypothesis as (1), and if the topological correlations vanish after the $\xi$th layer, then $\xi \geq \nu$. Therefore, (2) also means that $\xi \geq 3$. An important point in the theorem is that the Aboav parameter $a$ is a free parameter, for otherwise, we can have an even smaller $\nu$. (Indeed, if we restrict $a$ to be 1, then $\nu = 1$, and if we restrict $a$ to be 2, then $\nu = 2$, and if we do not put any restriction on $a$, then $\nu \geq 3$. ) Next we will explore some consequence of this theorem.

For a space filling Euclidean froth where the correlation length is minimal ($\xi = \nu = 3$) and the Aboav-Weaire's law is satisfied with $\Gamma_1 = (1-a)(6-n)$, eq.(53) implies $(1-a)\mu_2 + \langle (6-n)\Gamma_2 \rangle = -\mu_2$. A solution for this condition is given by $\Gamma_2 = (a-2)(6-n)$. Since we have $\xi = \nu = 3$, we can set $\Gamma_j = 0$ for $j \geq 3$ (from eq.(53) and (49) ). With these $\Gamma_j$, we can work out the solution for the number of cells per layer,

$$
\begin{aligned}
K_1 &= n \quad , \\
K_2 &= 12 + \mu_2 + (2-a)(n-6) \quad , \\
K_j &= \big(6 + (3-a)\mu_2\big)j - (5-2a)\,\mu_2 \quad \text{for } j \geq 3.
\end{aligned}
\tag{16}
$$

and the topological charge

$$
\begin{aligned}
Q_1 &= -\mu_2 + (1-a)(6-n) \quad , \\
Q_2 &= -(3-a)\mu_2 + (a-2)(6-n) \quad , \\
Q_j &= -(3-a)\mu_2 \qquad\qquad\qquad\qquad \text{for } j \geq 3.
\end{aligned}
\tag{17}
$$

The solution eq.(16) for the trajectories $K_j$ is qualitatively in agreement with the experimental data. The number of cells per layer versus $j$ follows a linear law with slope $6 + (3-a)\mu_2$ and intercept $-(5-2a)\mu_2$. Assuming the typical values $a \simeq 1$ and $\mu_2 \simeq 1.5$, we obtain 9 and -4.5 respectively for the typical slope and intercept. Note that a simple geometrical approach will suggest a linear growth of the number of cells of the perimeter of the shell cluster with a slope equal about $2\pi$, which differs from from the one we have found.

Equation (17) predicts that in SSI minimally correlated froths the topological charge is constant after the second shell. This is in contradiction with the experimental data which shows topological charges decreasing linearly with $j$. To resolve this contradiction with experimental data, we can consider the following scenario. If we take the fact that $Q_j \propto j$ for large $j$, then according to eq. 5 implies that $K_j \propto j^2$. But our experimental froth is Euclidean with $K_j \propto j$. Thus, the only possible way to get out of this contradiction with experimental data without invoking the results on non-SSI Euclidean froth is to conclude that the assumption that $\xi = \nu = 3$ in the above analysis is incorrect. Thus this conclusion, which insists on using the results of SSI Euclidean froth, leads one to suspect that topological correlation has a longer range: $\xi > 3$. However, this is actually misleading as we have no good reason to insist on the assumption that the real froth is SSI Euclidean. Indeed, we have enough evidence that the real froth is a non-SSI Euclidean froth, so that the topological correlation can vanish rather early, and still $K_j \propto j$ and $Q_j \propto j$ for large $j$. This point is also valid for computer generated froth. In the next section we show that the topological defects provide a mechanism which correctly gives the behaviour of the topological charge and the slope in the linear law of the trajectories $K_j$.

## 4    Effects of the non-inflatable inclusions

The ideas presented in the previous sections for shell-structured froths are still applicable when we have non-inflatable inclusions. In this section, we indicate the appropriate corrections. We use the symbols $n$, $K_j$, $Q_j$, $m_j$ and $\mu_2$ to indicate the quantities associated with the global froth (shell-skeleton plus topological inclusions) and we use the notation $K_j^{sk}$, $Q_j^{sk}$ and $m_j^{sk}$ for the quantities associated with the shell skeleton only. Finally, we use $K_j^d$ and $m_j^d$ to indicate the number of defective cells and their average number of sides. Note that $K_0^d = 0$ always (since any cell of the froth can be the germ cell). One has the simple relation

$$K_j = K_j^{sk} + K_j^d \qquad . \tag{18}$$

In the general case when non-inflatable inclusions are present, the relations previously obtained for the SSI froth are applicable to the quantities associated with the shell-skeleton. In particular relations (4) and (5) become

$$Q_j^{sk} = 6 - K_{j+1}^{sk} + K_j^{sk} \qquad , \tag{19}$$

and

$$K_j^{sk} = 6j - \sum_{i=0}^{j-1} Q_i^{sk} \qquad . \tag{20}$$

On the other hand, expression (3) for the total topological charge of the cluster delimited by shell $j$ remains unchanged

$$Q_j = 6 - V_j^+ + V_j^- \qquad , \tag{21}$$

here the quantities $Q$, $V^{\pm}$ are associated to the global froth (shell skeleton plus topological inclusions). This equation is the topological analog of the Gauss theorem of electrostatic. The charge (topological) inside a region of space is associated to the net flux (of edges) which cross the external surface(shell). In the absence of topological inclusion (SSI case) the flux of edges between two adjacent shells is uninterrupted (any edge outgoing from shell $j$ ends in shell $j+1$, i.e. $V_{j+1}^- = V_j^+$). In the general case, the topological inclusions trap the edges-flux and the previous identity must be modified as follows

$$V_{j+1}^- = V_j^+ - \eta_{j+1} K_{j+1}^d \qquad , \tag{22}$$

where $\eta_{j+1}$ is the average number of edges which are trapped by one defect in the layer $j+1$. One can easily verify that $V_j^- = K_j^{sk}$. By substituting eq.(22) and $V_j^-$ into eq.(21) one obtains

$$Q_j = 6 - K_{j+1}^{sk} + K_j^{sk} - \eta_{j+1} K_{j+1}^d \qquad , \tag{23}$$

which is the generalization of eq.(4). By substituting eq.(19) into (23) one gets

$$Q_j = Q_j^{sk} - \eta_{j+1} K_{j+1}^d \quad . \tag{24}$$

Therefore, the topological charge inside shell $j$ is equal to the charge associated with the shell-skeleton minus a contribution due to the defects attached to the external shell. This implies the important fact that the defects inside the cluster do not contribute to the total charge. Moreover, note that the defects attached to the external shell always decrease the topological charge in the shell respect to the value associated with the shell-skeleton.

## 4.1  First shell and Aboav-Weaire's law

The number of cells making the first layer around an $n$-sided central cell is $K_1 = n$. By using the sum rule (46) obtained in appendix A one can calculate the average topological charge inside the first shell

$$\langle Q_1 \rangle = \langle (6 - n) \rangle + \langle (6 - m_1) K_1 \rangle = \langle (6 - n) n \rangle = -\mu_2 \quad . \tag{25}$$

From this relation and eq.(24) follows that $\langle Q_1^{sk} \rangle = -\mu_2 + \langle \eta_2 K_2^d \rangle$.

As in the SSI case, one can introduce the fluctuations of the topological charge around its average value

$$\Gamma_j = Q_j - \langle Q_j \rangle \quad . \tag{26}$$

In term of these fluctuations and using eq.(25), the charge contained in the first layer can be written as

$$Q_1 = (6 - n) + (6 - m_1) K_1 = -\mu_2 + \Gamma_1 \quad . \tag{27}$$

This equation gives a relation between the number of sides ($n$) of a given cell and the average number of sides ($m_1$) of the cells surrounding this $n$ sided cell. In the particular case that the fluctuation is linear in $n$, $\quad \Gamma_1 = (1 - a)(6 - n)$, we arrive at the Aboav-Weaire's law (i.e. $n m_1 = (6 - a) n + \mu_2 + 6a$ [15]). The Aboav's coefficient $a$ can be interpreted (see §2.1) as the factor that represents the screening of the central charge due to the surrounding cells. By following this interpretation one can associate $a = 1$ to a total screening, and $a = 0$ to the absence of screening. A typical value for this coefficient in natural cellular structures (soap froth, alumina cuts, etc.) is $a \simeq 1.2$, values around $0.6$ are characteristic of Voronoi froths constructed from Poissonian points, whereas random network generated by performing T1 transformations on regular lattice have negative Aboav's coefficients $a \simeq -1$.

## 4.2  Generic $j$-shell

The average value of the topological charge inside the layer $j$ can be calculated by using the sum rule (46) (i.e. $\langle (6 - m_j) K_j \rangle = \langle (6 - n) K_j \rangle$, see Appendix A) and the expression (20) for $Q_j^{sk}$, to obtain

$$\begin{aligned} \langle (6 - m_j) K_j \rangle &= \langle (6 - n) K_j^d \rangle + \langle (6 - n) K_j^{sk} \rangle \\ &= \langle (6 - n) K_j^d \rangle - \sum_{i=0}^{j-1} \langle (6 - n) Q_i^{sk} \rangle \end{aligned}$$

$$
\begin{aligned}
&= \langle (6-n)K_j^d \rangle - \sum_{i=0}^{j-1} \Big( \langle (6-n)Q_i \rangle + \langle (6-n)\eta_{i+1}K_{i+1}^d \rangle \Big) \\
&= \langle (6-n)K_j^d \rangle - \sum_{i=0}^{j-1} \Big( \langle (6-n)\Gamma_i \rangle + \langle (6-n)\eta_{i+1}K_{i+1}^d \rangle \Big) \quad . \qquad (28)
\end{aligned}
$$

The average total charge of the cluster inside shell $j$ is the sum over the charge of individual layers. From eq.(28) follows

$$
\langle Q_j \rangle = \sum_{i=0}^{j-1} \Big( \langle (6-n)K_{i+1}^d \rangle - (j-i)\big( \langle (6-n)\Gamma_i \rangle + \langle (6-n)\eta_{i+1}K_{i+1}^d \rangle \big) \Big) \quad . \qquad (29)
$$

In this expression, the $i=0$ term is $\langle (6-n)K_1^d \rangle - j(\mu_2 + \langle (6-n)\eta_1 K_1^d \rangle)$, since $\Gamma_0 = 6-n$. From eq.(20), using eq.(29) and the definition of $\Gamma_i$, the number of cells of the shell-skeleton in the layer $j$ is

$$
\begin{aligned}
K_j^{sk} &= 6j - \sum_{i=0}^{j-1} Q_i^{sk} \\
&= 6j - \sum_{i=0}^{j-1} \Big( Q_i + \eta_{i+1}K_{i+1}^d \Big) \\
&= 6j - \sum_{i=0}^{j-1} \Big( \Gamma_i + \langle Q_i \rangle + \eta_{i+1}K_{i+1}^d \Big) \\
&= 6j - \sum_{i=0}^{j-1} \Big( \Gamma_i + \eta_{i+1}K_{i+1}^d + (j-i)\langle (6-n)K_i^d \rangle \Big) \\
&\quad + \sum_{i=0}^{j-2} \frac{(j-i)(j-i-1)}{2} \Big( \langle (6-n)\Gamma_i \rangle + \langle (6-n)\eta_{i+1}K_{i+1}^d \rangle \Big)
\end{aligned}
$$
$$(30)$$

This is the generalization of eq.(15) which takes into account non-inflatable inclusions. These relations have the beauty of being exact and the privilege of being useless for predicting properties of real space-filling cellular systems. In order to compare with experiments, we must use some simple physical approximations that will provide predictions for the asymptotic behaviours of $K_j$, $Q_j$, and the percentage of defects.

## 5    Euclidean froths and non SSI inclusions

In the absence of defects and fluctuations eq.(30) implies that the number of cells in each layer grows very fast ($K_j \propto j^2$). In principle, such a fast growth is realizable in the $2D$ plane by increasing continuously the roughness of the cluster-surface enlarging the available space for new cells in the layer. In practice, after some layers the rough surface of the cluster starts to self-intersect, thereby generating non SSI inclusions (topological defects in the shell structure). These inclusions of defective cells provide a way to smooth the shell surface and to keep the froth

Euclidean as a consequence. Therefore, the defects play a very important role in the froth-organization.

Soap froths and computer generated random cellular systems show asymptotically a linear increment in the number of cells per layer with the distance from the germ cell. Let's therefore consider an Euclidean froth where $K_j \propto j$ for any $j \geq \nu$. In contrast to the SSI case, when we have defects, this condition of linear growth of $K_j$ does not introduce any constraint on the fluctuations $\Gamma_j$. Indeed, the number of defects $K_j^d$ is a free parameter that the system can adjust in order to control the fluctuations in the topological charge and simultaneously keep the froth Euclidean. Analogously, the defects can play an important role in the reduction of topological correlations. In Section 3 we found that an SSI Euclidean froth which satisfies the Aboav-Weaire's law must be correlated at least between third neighbours. The defects enlarge the freedom for the construction of the cellular system around the germ cell by removing the constraints on the fluctuations. (Without these constraints on the fluctuations, one can construct an Euclidean froth which satisfy the Aboav-Weaire's law and which is correlated only between first neighbours.) Therefore the presence of defects can strongly reduce the correlations.

Consider a froth which is minimally correlated and compatible with an Aboav-Weaire's law with a free $a$ parameter. From eq.(49) (see Appendix B) it follows that such a froth can be uncorrelated after the first neighbours ($\xi \geq 1$). (Indeed, for the topological gas, $\xi = 0$ and this corresponds to $a = -\mu_2/6$ ) In this case, one gets using eq.(30) and $\Gamma_1 = (1 - a)(6 - n)$,

$$K_2 = 12 + \mu_2 + (2 - a)(n - 6) - \eta_1 K_1^d + (1 - \eta_2) K_2^d + \langle (n - 6)(1 - \eta_1) K_1^d \rangle \qquad . \qquad (31)$$

Transcurating the contribution from the defects in the first layer ($K_1^d \ll 1$) and fixing $\eta_2 = 1$, eq.(31) become

$$K_2 \simeq 12 + \mu_2 + (2 - a)(n - 6) \qquad . \qquad (32)$$

Equation (49) gives

$$\Gamma_2 = \left(1 - a + \frac{(2 - a)^2 \mu_2}{12 + \mu_2}\right)(6 - n) \qquad . \qquad (33)$$

By using eq.(30) one has

$$K_3 \simeq 18 + (4 - a)\mu_2 + \left(3 - 2a + \frac{(2 - a)^2 \mu_2}{12 + \mu_2}\right)(n - 6) - K_2^d \qquad (34)$$

where we used the same approximations as for eq.(32) plus the hypothesis $\eta_3 = 1$.

These relations are derived using the assumptions that $K_1^d \ll 1$, $\xi = 2$ and $\eta_2 = \eta_3 = 1$. The first condition $K_1^d \ll 1$ is quite reasonable as experimentally in real or computer generated froth, the first few shells have small amount of defects. The second condition is a working condition on the conventional wisdom that froth has rather short range correlations. In any case, this is a working hypothesis for deriving eq.32, 33 and 34, which are all the equations needed for comparison with experimental data in the next section. As for the third condition $\eta_2 = \eta_3 = 1$, this is actually a reasonable assumption as $\eta_{j+1}$ is the average number of edges which are trapped by one defect in the layer $j + 1$, and many defects contribute in general only one extra edge to the correction in the defining eq.22.

## 5.1 Asymptotic behaviour in Euclidean froths

In an Euclidean $2D$ froth, the space-filling condition implies that in the asymptotic limit the number of cells per layer grows linearly with the topological distance and consequently at each layer this number is incremented by a constant amount. In a froth where the cells are uncorrelated after a given topological distance $\xi$ this rate of increment must be a constant parameter characteristic of the whole froth and independent of the particular central cell. The dependence of $K_j$ on the number of sides $(n)$ of the central cell can only be an additive constant (i.e. $K_j = Cj + B(n)$). From a geometrical point of view this additive quantity is associated to the size of the central core made by the first shell and its neighbours. In a system that is uncorrelated after the first shell, the length of the perimeter at shell $j$ is given by the perimeter of this core $(B(n))$ plus a term linearly dependent by the distance $(Cj)$. The generalized Aboav-Weaire's relation (eq.(46) in Appendix A), gives

$$\langle (6 - m_j)K_j \rangle = \langle (6 - n)B(n) \rangle \qquad \text{for } j \geq \xi. \tag{35}$$

By assuming $\xi = 2$ and using eq.(32) we obtain

$$\langle (6 - m_j)K_j \rangle = \langle (6 - n)K_2 \rangle \simeq -(2 - a)\mu_2 \qquad \text{for } j \geq 2. \tag{36}$$

Equation (36) states that the topological charge contained inside any layer is a constant quantity equal to $-(2 - a)\mu_2$. It follows that the total topological charge inside shell $j$ decreases linearly with $j$

$$\langle Q_j \rangle = -(2 - a)\mu_2 j + const. \qquad \text{for } j \gg 1. \tag{37}$$

Note that the total topological charge in the froth must be lower than 12[31]. Indeed, a froth with charge $Q = 12$ is a closed cellular system which is tiling a surface topologically equivalent to a sphere. Therefore, from eq.(37) it follows that the Aboav coefficients must be smaller or equal than 2.

In an Euclidean $2D$ froth, the number of cells per layer grows linearly with slope $C$ asymptotically. Let us suppose that the percentage of defects with respect to the total number $K_j$ of cells in the layers is a constant $\Lambda$ independent of the topological distance in this limit. Then the number of cells in the shell-skeleton must also grow linearly with $j$ (since $\langle K_j^{sk} \rangle = (1 - \Lambda)\langle K_j \rangle$ and $\langle K_j \rangle$ is linear in $j$). Equation (20) indicates that $\langle K_j^{sk} \rangle$ can be linear if the average topological charge associated with the shell skeleton $\langle Q_j^{sk} \rangle$ is constant. This implies (eq.(24) )

$$
\begin{aligned}
\langle (6 - m_j)K_j \rangle &= \langle Q_j - Q_{j-1} \rangle \\
&= \langle Q_j^{sk} - Q_{j-1}^{sk} \rangle - \langle \eta_{j+1} K_{j+1}^d - \eta_j K_j^d \rangle \\
&= \left( \eta_j K_j^d \rangle - \langle \eta_{j+1} K_{j+1}^d \rangle \right) \qquad \text{for } j \gg 1.
\end{aligned}
\tag{38}
$$

By supposing the parameter $\eta_j = \eta$ independent of $j$ for large $j$, using eq.(38) we get

$$\langle (6 - m_j)K_j \rangle \simeq -\eta\Lambda\left( \langle K_{j+1} \rangle - \langle K_j \rangle \right) = -\eta\Lambda C \qquad \text{for } j \gg 1, \tag{39}$$

which predicts that the charge decreases linearly with $j$ with a decrement of $\eta\Lambda C$ per layer. From eq.(39) and using the expression for $m_j$ obtained in Appendix B (eq.(47)) valid for layers of cells uncorrelated with the germ cell (in present case for $j > 2$), we obtain

$$m_j = m_j^{un} = 6 - \frac{\langle (6-n)K_j \rangle}{\langle K_j \rangle} \simeq 6 + \frac{\eta \Lambda C}{Cj + \langle B \rangle} \simeq 6 + \frac{\eta \Lambda}{j} \ . \tag{40}$$

Supposing that $\langle K_j \rangle$ grows linearly from the second shell (i.e. fixing $\nu = 1$), from eq.(31) and (34) we obtain the slope

$$C \simeq \langle K_3 \rangle - \langle K_2 \rangle \simeq 6 + (3-a)\mu_2 - \langle K_2^d \rangle \ . \tag{41}$$

By comparison between eq.(36) and (39) and using expression (41) we obtain an expression for the percentage of defects

$$\eta \Lambda \simeq \frac{(2-a)\mu_2}{6 + (3-a)\mu_2 - \langle K_2^d \rangle} \simeq \frac{(2-a)\mu_2}{6 + (3-a)\mu_2} \ , \tag{42}$$

where we assumed $\langle K_2^d \rangle \ll 6 + (3-a)\mu_2$. From eq.(42) one notes that a froth can be free of defects only if $\mu_2 = 0$ or $a = 2$. The first case ($\mu_2 = 0$) corresponds to the hexagonal lattice which is SSI and therefore free of defects. The second case ($a = 2$) corresponds to a froth where the Aboav parameter takes the maximum allowed value for an Euclidean froth. This is a very peculiar froth which has constant topological charge in the shell clusters and does not need defects to fill the plane. So far such a froth has not been observed. Note that $a = 2$ is a critical value since froths with $a > 2$ are closed elliptic systems. That might suggest that the value $a = 2$ is an upper limit which cannot be reached for Euclidean froths.

In an earlier paper [22], we have used the shell model to test a generalization of the Aboav-Weaire's law to shells beyond the first. These analysis reveal a universal topological relation on the average number $m_j$ of sides per cell to the number of cells $K_j$ in the $j$-th layer of a given center cell with $n$ sides. A plot of $m_j K_j$ vs $K_j$ shows a slope of $(6-a)$ for $j = 1$(Aboav-Weaire's law) and a slope of 6 for $j \geq 2$ for all samples. The results are universal for soap froths in the scaling state with different preparations, different times, and different temperatures. With the small $\eta \Lambda$ approximation and eq.40, we can now provide a quantitative explanation to the experimental results. The average number of neighbors $m_j$ in layer of cells uncorrelated with the central cell ($j \geq \xi$) is given by eq.(47). Multiplying this expression by $K_j$ we have

$$m_j K_j = m_j^{un} K_j = \left( 6 - \frac{\langle (6-n)K_j \rangle}{\langle K_j \rangle} \right) K_j \qquad (j \geq \xi), \tag{43}$$

In the asymptotic limit, using the approximations introduced above, we obtain

$$m_j K_j \simeq 6K_j + (2-a)\mu_2 \qquad (j \geq \xi), \tag{44}$$

which is the extension of the Aboav-Weaire's law to higher shell numbers.

# 6 Experimental results and comparison with the theoretical predictions

The soap froth chamber consists of two 1.6 cm thick retangular plexiglass plates separated by a 0.16cm thick spacer. The effective working area for the froth is 26.7cm×36.8cm which can be filled with more than twenty thousand bubbles as the starting condition. Soap bubbles of different sizes are pumped into the chamber through an inlet to create a relatively random froth. The chamber is filled with excess bubbles such that excess fluid can be forced out of the chamber through another outlet after the soap froth has been drained for a few minutes by setting the chamber vertically. Thus the froth has a negligible volume fraction of liquid to air. The chamber is then sealed by closing the inlet and outlet and placed horizontally. A dark field method is used for viewing from above and a high resolution digital CCD camera of $1037 \times 1344$ pixels is used to capture the images at different stages of the evolution of the froth. The experiment starts with about 20,000 bubbles and is about 10,000 in the scaling state defined by the stationary distribution of sides after about 4 hours. However, the topological charge becomes stationary at a much later time with about 4,000 bubbles. Runs with similar initial condition have been recorded and only data with stationary topological charge will be reported in this paper.

In order to compare the physical froth with computer generated ones, we have selected two examples of purely geometric constructions. The first one is the Voronoi construction based on random Poissonian points. The second one is based on the Voronoi construction generated by introducing small perturbations to the triangular lattice so that we mimick the T1 transformation in real froth. We label the physical froth by (S), the random Voronoi froth by (V), and the perturbed one by (P). These two geometric constructions are two simple examples of the random froth and slightly ordered froth. The number of cells for these systems are: 3206 for (S), 3783 for (V) and 9634 for (P). The values of $\mu_2$ are : 1.33 (S), 1.76 (V), 1.51 (P). They all obey approximately the Aboav-Weaire's law with coefficient $a$ equal respectively to $1.27\pm0.05$ (S), $0.69 \pm 0.03$ (V), $0.95 \pm 0.07$ (P).

In Fig.(2) the number of cells per layer $K_j$ is shown as function of the distance $j$ for soap, Voronoi construction on set of random points in the plane and on perturbed triangular lattice. For all systems this number increases linearly with the distance with slopes respectively equal to $9.45\pm0.1$ (S), $11.0\pm0.2$ (V), and $9.91\pm0.08$ (P). These linear behaviours indicate that these froth are Euclidean. Using the expressions for the minimally correlated Euclidean froths obtained in the previous paragraph, one can predict the slopes $8.3\pm0.3$(S), $10.1\pm0.4$(V) and $9.1\pm0.7$(P).

The ratios $\Lambda$ of the defective cells respect to the total number of cells in a layer are very small in the second layer ($0.0173$(S), $0.05$(V) and $0.05$(P)), then increase until about the $15^{th}$ layer, to stabilize finally at an asymptotic value equal to $0.10\pm0.01$ (S), $0.15\pm0.01$ (V), and $0.13\pm0.01$ (P) respectively. Using the expression (42) one can predict for $\eta\Lambda$ the values $0.136$ (S), $0.23$ (V) and $0.174$ (P) respectively. We measured independently the value of $\eta$ which gives $1.1 \pm 0.2$ (S), $1.3\pm0.1$ (V), and $1.2\pm0.1$ (P) respectively. We deduce that the value of defect concentration $\Lambda$ using these values of $\eta$ and Eq.(42) are $\Lambda = 0.11 \pm 0.03$ (S), $0.17\pm0.02$ (V), and $0.15\pm0.03$ (P) respectively. This compares reasonably well with the measured values of $\Lambda$.

Fig.(4) shows the total average charge $\langle Q \rangle/\mu_2$ v.s. $j$. This quantity decreases with the distance, with an almost linear behaviour for $j > 2$. The measured slopes are respectively -0.7$\pm$0.1 (S), -1.6$\pm$0.1 (V), and -1.35$\pm$0.05 (P). Using equation (37) one predicts -0.73$\pm$0.03 (S), -1.31$\pm$0.05 (V), and -1.05$\pm$0.07 (P).

The topological charge contained inside the cluster delimited by the generic shell $j$ is an important physical parameter. In the previous paragraph we showed that the quantity associated with the shell-structured skeleton $Q_j^{sk}$ should be constant for Euclidean froths. In Fig.(4) $\langle Q^{sk} \rangle/\mu_2$ is plotted (open symbols) in function of $j$ for the different froths studied. One can note that in all the systems analysed the quantity $\langle Q^{sk} \rangle / \mu_2$ saturates at a value equal about to -2.0±0.5 (S), -2.0±0.2 (V), and -1.8±0.2 (P). The saturation values should be similar for all three cases as $\langle Q^{sk} \rangle / \mu_2$ is a characteristic of the SSI skeleton.

# 7   Conclusions

We have studied the froth as organized in concentric layers of cells around a given central cell. Exact expressions for the number of cells in each layer, for the topological charge inside the shell cluster and for the average number of neighbors per cell in a given layer have been obtained. These topological properties of the shell-structure have been studied for the special class of SSI froths and for the general case where non-SSI defects are present. It turns out that the defects play a very important role in the organization of the froth structure. The defects enlarge the freedom for the construction of the cellular system around the germ cell by removing the constraints on the topological charge associated with Euclidean froths. With the relaxation of the constraints by the introduction of defects, we find a solution of space filling Euclidean froth which has only nearest neigbhors correlations. In this solution, we have calculated approximate asymptotic expressions for $K_j$, $Q_j$ and $m_j$ which satisfy the Aboav relation and which are correlated only between first neighbors. Moreover, we have evaluated the average number of defects per layer. These expressions are free of adjustable parameters and describe well the behaviors of measurable properties of real froths and cellular patterns.

Experimentally we find that soap froth in the steady state has an average number of cells per layer which grows linearly with the topological distance. The rate of growth is about 9. Slopes around 10 have been found for Voronoi froths. These slopes are considerably bigger than $2\pi$ value suggested by simple geometrical consideration for the ratio between the perimeter of the shell-cluster and its radius. Moreover, we found that in soap and Voronoi froths the topological charge of the shell-cluster is always negative and decreases linearly with the size of the cluster. In particular, we observed that the slope is proportional to $-\mu_2$ and the coefficient of proportionality is smaller than 1 in soap froths and bigger than 1 in computer generated froths. The number of defects per layer have been also measured. Soap froths have in the asymptotic limit a percentage of defects around 10%, whereas bigger amounts have been found for computer generated froths.

Our theory on the asymptotic behaviors is in good qualitative agreement with experiments. We correctly predict the linear growth of the number of cells per layer with a slope around 9. We demonstrate the linear decrement in the topological charge of the shell-cluster with a slope above the line defined by $-\mu_2$ in soap froths and with a slope below the line defined by $-\mu_2$ in Voronoi froths. We also predict percentages of defects per layer which are close to the experimental values and are smaller in soap froths and bigger in Voronoi froths.

On the other hand, the quantitative agreement between the approximated predictions and the experimental data is not perfect. In the present paper we have obtained exact relations for the topological properties of the shell structure, but the predictions have been formulated under strong assumptions which simplify the exact results into expressions with no adjustable parameters. The partial disagreement between the approximated predictions and the experimental data might indicate that the assumption utilized are too strong or incorrect. Therefore there is still room for improvement.

**Acknowledgements**

We acknowledge discussion with N. Rivier. T. Aste acknowledges support from the Hong Kong University of Science and Technology and the partial support from EU, HCM Program "Physics of Foams" CHRXCT940542 and by the TMR contract ERBFMBICT950380. K.Y. Szeto acknowledges support from the Hong Kong Telecom Institute of Information Technology. W.Y. Tam acknowledges support from the UPGC Research Infrastructure Grant (RI93/94.EG05) of HKUST.

# A    Generalization of the Weaire's sum rule

Consider an $i$-cell of the froth with $n_i$ sides and the sum over the number of sides of the set of cells at a topological distance $j$ from the $i$-cell. Such a sum is equal to $m_j(n_i)K_j(n_i)$ where $m_j(n_i)$ denotes the average number of sides per cell in the layer distant $j$ from the $i$-cell and $K_j(n_i)$ denotes the number of cells of this layer. Let us now sum this quantity over all the cells of the system: $\sum_i m_j(n_i)K_j(n_i)$. In this sum the number of edges of each $i$-cell of the froth is counted a number of time equal to the number of cells at distance $j$ from this cell (e.g. the number of sides of the generic $i$-cell is counted $K_j(n_i)$ times and contribute to the sum as $n_i K_j(n_i)$. It follows that we have the identity

$$\sum_i m_j(n_i)K_j(n_i) = \sum_i n_i K_j(n_i) \quad . \tag{45}$$

We can express this identity in term of the averages

$$\langle m_j K_j \rangle = \langle n K_j \rangle \tag{46}$$

where $\langle (...) \rangle$ indicates the average over the cell-sides distribution: $\langle (...) \rangle = \sum_n p(n)(...)$ with $p(n)$ probability of an $n$-sided cell.

Equation (46) is the generalization to the layer $j$ of the Weaire's sum rule which is valid for the first layer (i.e $\langle m_1 n \rangle = \langle n^2 \rangle$). For an arbitrary large system (arbitrary small boundary effects) relation (46) is exact and is valid also in the presence of non shell-reducible topological inclusions. One should note that, in real finite systems the effect of the boundary could be dramatically important.

# B    Correlation and fluctuations

A froth is uncorrelated after a given topological distance $\xi$ if and only if for two cells respectively distant $j > \xi$, the probability $C_j(n,m)$ to have one with $n$ sides and the other with $m$ sides factories $C_j(n,m) = s_j(n)s_j(m)$. It can be easily proved that in a froth where the cells are uncorrelated after the distance $\xi$, the average number of sides per cell $(m_j^{un})$ in a layer $j > \xi$ must be independent of the number of sides of the central cell. This is the physical consequence of the absence of correlations between the central cell and the cells in the layer $j$. This therefore means $\langle m_j^{un} K_j \rangle = m_j^{un} \langle K_j \rangle$. By using relation (46), we have

$$m_j = m_j^{un} = 6 - \frac{\langle (6-n)K_j \rangle}{\langle K_j \rangle} = \frac{\langle n K_j \rangle}{\langle K_j \rangle} \quad .(\text{for } j > \xi). \tag{47}$$

The topological charge inside layer $j$ is

$$Q_j - Q_{j-1} = (6 - m_j)K_j = \langle (6-n)K_j \rangle + \Gamma_j - \Gamma_{j-1} \qquad , \qquad (48)$$

The second equality comes from eq.(A2). For $j > \xi$, we put eq.(B1) into (B2) to get for the uncorrelated case the relation

$$\Gamma_j = \Gamma_{j-1} + \langle (6-n)K_j \rangle \left( \frac{K_j}{\langle K_j \rangle} - 1 \right) \qquad (\text{for } j > \xi). \qquad (49)$$

This is a recursive equation which gives the fluctuations $\Gamma_j$. Therefore, in uncorrelated froths the topological charge fluctuation are determined in term of the other statistical properties of the shell system.

Consider for example a froth which is completely uncorrelated (topological gas, $\xi = 0$). From eq.(49) one gets

$$\Gamma_1 = \left( 1 + \frac{\mu_2}{6} \right)(6 - n) \qquad , \qquad (50)$$

where we used the identities $K_1 = n$ and $\Gamma_0 = 6 - n$. This relation gives the Aboav-Weaire's law ($\Gamma_1 = (1-a)(6-n)$, see in the text) with coefficient $a = -\mu_2/6$.

# C   Proof for the correlation theorem of Euclidean SSI froth

To prove this theorem, we first note that eq.(5) implies

$$\langle K_j \rangle = 6j - \sum_{i=0}^{j-1} \langle Q_i \rangle \qquad . \qquad (51)$$

In order to fill the two-dimensional Euclidean space, there must be a minimum $\nu$ such that $\langle K_j \rangle \propto j$ for $j \geq \nu$. This implies that the cellular system must constrain the average topological charge inside a shell to be independent on the shell-size $j$. Such a constraint forces the average charge inside the layer $j$ to be equal to zero for $j \geq \nu \geq 1$, (see eq.13)

$$\langle Q_j \rangle - \langle Q_{j-1} \rangle = \langle (6-m_j)K_j \rangle = -\sum_{i=0}^{j-1} \langle (6-n)\Gamma_i \rangle = -\mu_2 - \sum_{i=1}^{j-1} \langle (6-n)\Gamma_i \rangle = 0 \qquad . \qquad (52)$$

Consequently, one gets the following two conditions on the fluctuations $\Gamma_i$

$$\sum_{i=1}^{\nu-1} \langle (6-n)\Gamma_i \rangle = -\mu_2 \qquad \text{for } \nu \geq 1 \qquad , \qquad (53)$$

and

$$\langle (6-n)\Gamma_j \rangle = 0 \qquad \text{for } j \geq \nu \qquad . \qquad (54)$$

In appendix B we show that in a system where the topological correlations vanish after the $\xi^{th}$ layer, the fluctuations must satisfy the following relation for $j > \xi$ (eq.(B3)),

$$\langle (6-n)\Gamma_j \rangle = \langle (6-n)\Gamma_{j-1} \rangle + \frac{\langle (6-m_j)K_j \rangle^2}{\langle K_j \rangle} \qquad . \qquad (55)$$

If one supposes $\nu > \xi$ and if we set $j = \nu$ in eq.(55), equations (52) and (54) yield $\langle (6-n)\Gamma_{\nu-1}\rangle = 0$ which is in contradiction with the definition of $\nu$ (for $\mu_2 \neq 0$). That implies $\nu \leq \xi$. An immediate consequence is that in a random SSI froth where the Aboav-Weaire's law is satisfied with an arbitrary $a$ (i.e. $\Gamma_1 = (1-a)(6-n)$, for some general $a$), the cells must be correlated at least between third neighbours. Indeed, $\nu = 1$ in eq.(54) implies $(1-a)\mu_2 = 0$, or $a = 1$ if $\mu_2 \neq 0$, whereas $\nu = 2$ in eq.(53) implies $(1-a)\mu_2 = -\mu_2$, or $a = 2$ if $\mu_2 \neq 0$. Both cases restrict $a$ to have special values for $\mu_2 \neq 0$. It follows that if $a$ is to remain a free parameter of the froth, $\nu \geq 3$. Since we have shown that $\xi \geq \nu$, therefore $\xi \geq 3$.

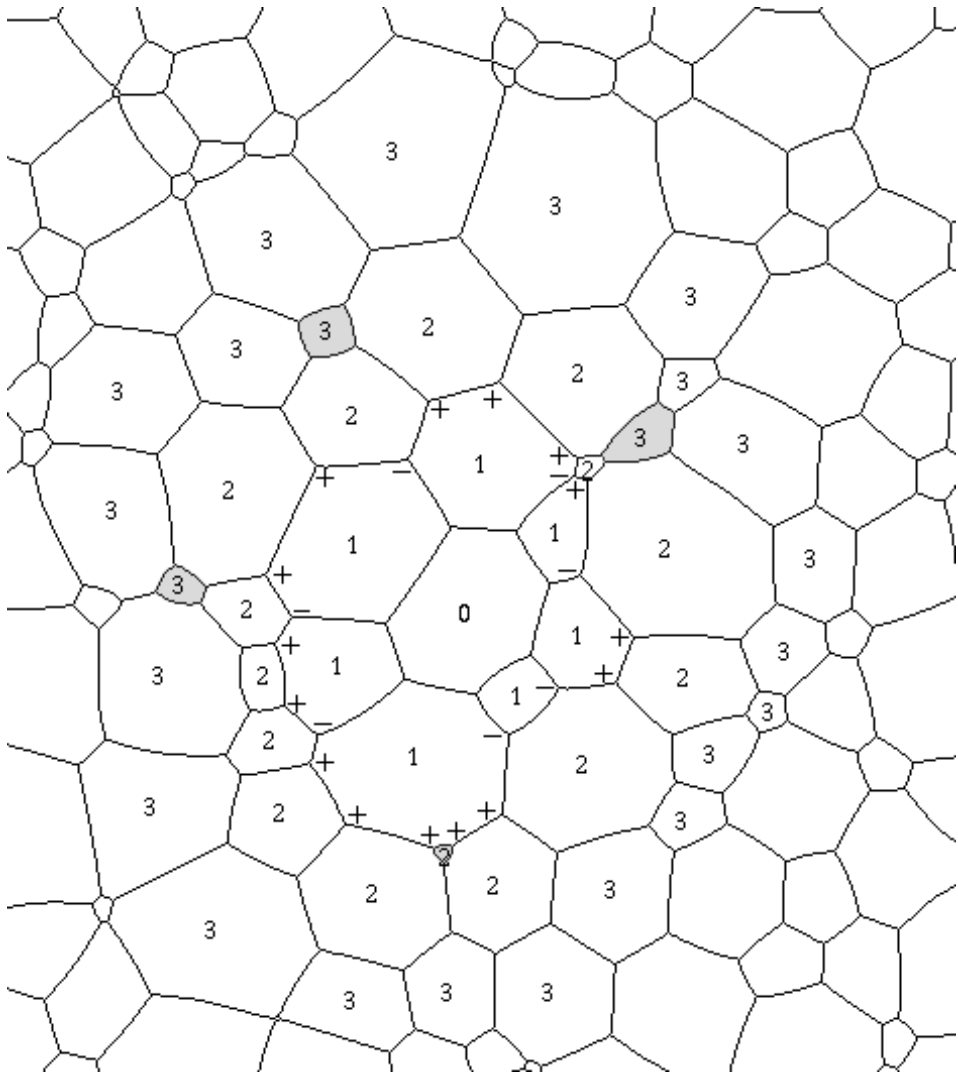

Figure 1: Shell structure and defects in trivalent two-dimensional froth, the defects are shaded and the number denotes the topological distance from the center cell labeled $O$, which is a deformed heptagon.

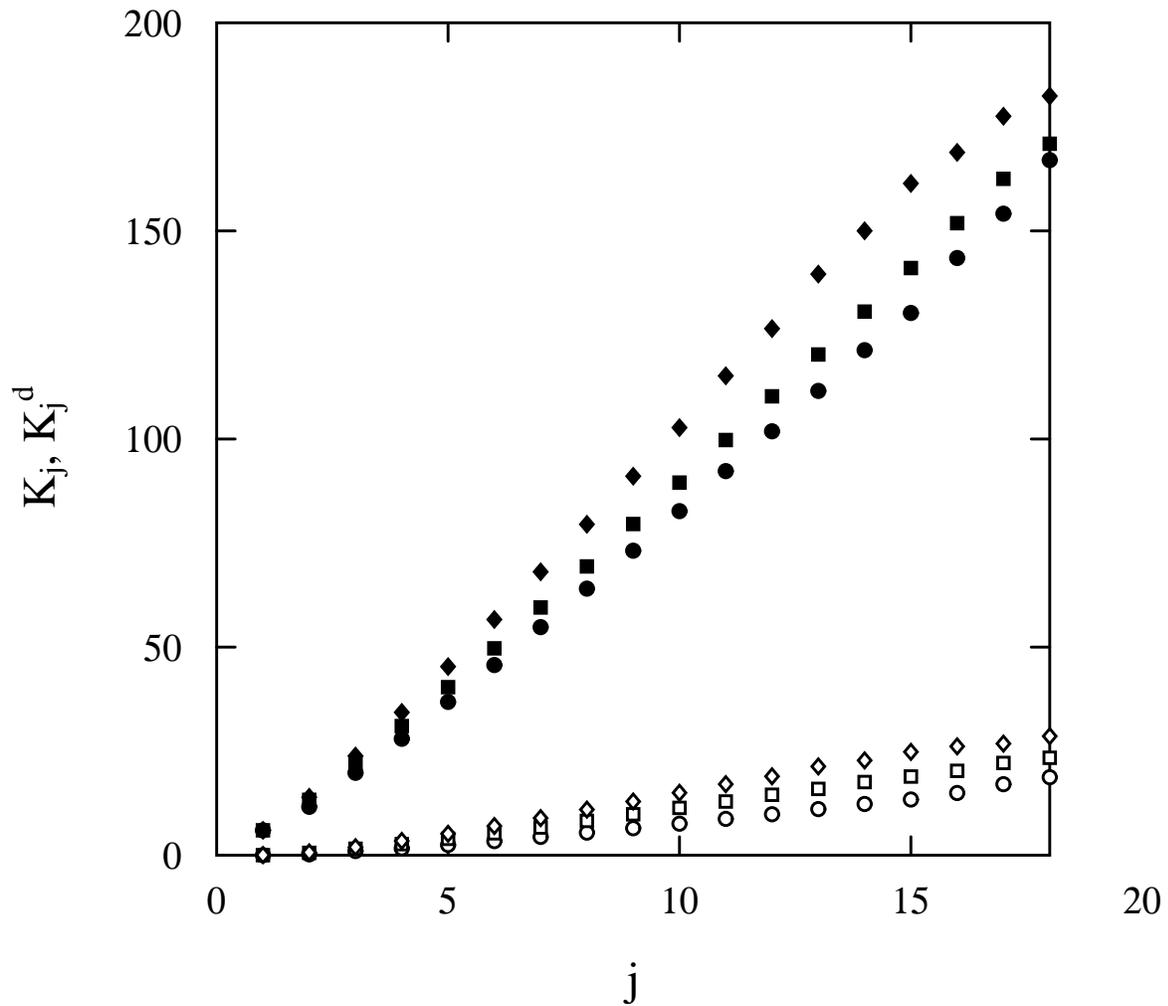

Figure 2: Total number of cells per layer $K_j$ in the $j$ shell vs shell number $j$ for soap (filled circle), random Voronoi construction (filled diamond) and Voronoi construction from perturbed triangular lattice (filled square). Total number of defects $K_j^d$ in the $j$-th shell vs $j$ for soap (open circle), random Voronoi construction (open diamond) and Voronoi construction from perturbed triangular lattice (open square).

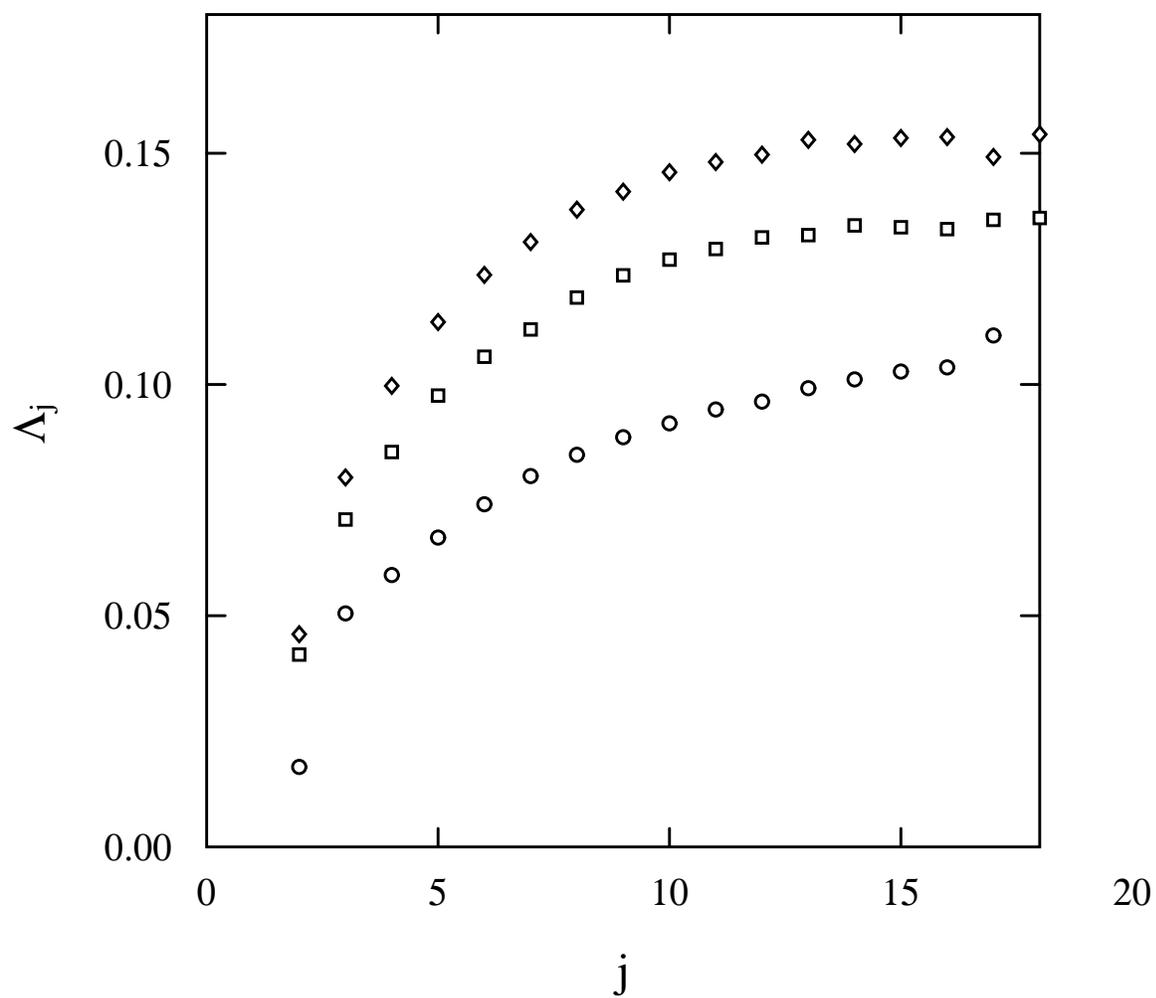

Figure 3: Defect concentration $\Lambda_j$ vs shell number $j$ for soap (open circle), random Voronoi construction (open diamond) and Voronoi construction from perturbed triangular lattice (open square).

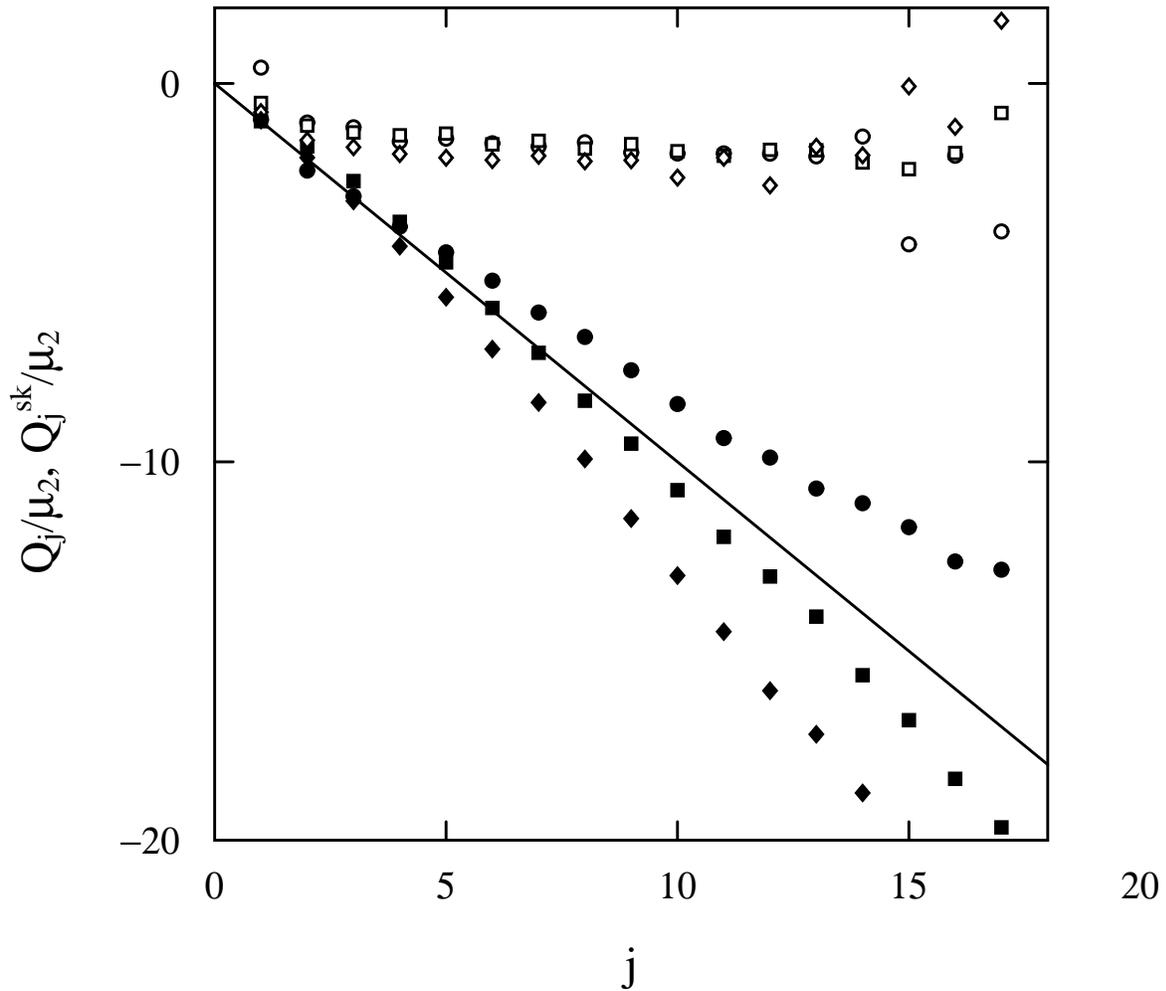

Figure 4: Normalized cluster topological charge $Q_j/\mu_2$ vs shell number $j$ for soap (filled circle), random Voronoi construction (filled diamond) and Voronoi construction from perturbed triangular lattice (filled square). The straight line indicate the slope -1. Soap and computer generated froths are in the opposite sides of this line. Normalized cluster topological charge of the skeleton, $Q_j^{sk}/\mu_2$ vs shell number $j$ for soap (open circle), random Voronoi construction (open diamond) and Voronoi construction from perturbed triangular lattice (open square). Note that the normalized topological charge for the skeleton saturated at about the same value for all cases.